\documentclass{PoS}

\usepackage{graphicx}
\usepackage{pgfplots}
\usepackage{algorithmic}
\usetikzlibrary{external}
\pgfplotsset{width=10cm, height=7cm } %,compat=1.3}
\usepackage{xcolor}
\renewcommand{\vec}{\mathbf}
\newcommand{\inner}[2]{\langle #1, #2 \rangle}
\usepackage{url}

\title{GPUs: An Oasis in the Supercomputing Desert}

\ShortTitle{GPUs: An Oasis in the Supercomputing Desert}

\author{\speaker{Waseem Kamleh} \\%
        University of Adelaide, Australia\\
        E-mail: \email{waseem.kamleh@adelaide.edu.au}}

\abstract{A novel metric is introduced to compare the supercomputing
  resources available to academic researchers on a national
  basis. Data from the supercomputing Top 500 and the top 500
  universities in the Academic Ranking of World Universities (ARWU)
  are combined to form the proposed ``500/500'' score for a given
  country. Australia scores poorly in the 500/500 metric when compared
  with other countries with a similar ARWU ranking, an indication that
  HPC-based researchers in Australia are at a relative disadvantage
  with respect to their overseas competitors. For HPC problems where
  single precision is sufficient, commodity GPUs provide a
  cost-effective means of quenching the computational thirst of
  otherwise parched Lattice practitioners traversing the Australian
  supercomputing desert. We explore some of the more difficult terrain
  in single precision territory, finding that BiCGStab is unreliable
  in single precision at large lattice sizes. We test the CGNE and CGNR forms of the
  conjugate gradient method on the normal equations. Both CGNE and a modified form of CGNR (with restarts) provide
  reliable convergence for quark propagator calculations in single precision. }

\FullConference{The 30th International Symposium on Lattice Field Theory\\
		 June 24 -- 29,  2012\\
		 Cairns, Australia}

\begin{document}

\section{The 500/500 Metric for Academic HPC Resources}

%\noindent\rule{\textwidth}{0.5cm}
%\noindent\rule{430pt}{0.5cm}\\
%\the\textwidth\\
%\the\textheight\\
%\the\paperwidth\\
%\the\paperheight

Lattice QCD has traditionally and continues to be one of the most
computationally demanding research fields within quantitative
science. Progress in Lattice QCD has closely tracked advances in high
performance computing (HPC). It is unsurprising then that the semi-annual supercomputing Top 500
list\footnote{\url{http://top500.org/}} is closely watched by many researchers within
lattice QCD. The Top 500 provides a straightforward answer to those
wanting to know which country has the biggest and the best
machines. It is arguable that such a simple comparison is not always
the most relevant. In certain circumstances,
% such as when writing grant applications for supercomputing time, 
it may be more pertinent
to ask a different question: \emph{How much supercomputing access do I
  have relative to my competitors overseas?}

In an attempt to provide an answer, our starting point is the Academic
Ranking of World Universities (ARWU) list compiled by Shanghai Jiao
Tong University, China, also known as the Shanghai
Ranking\footnote{\url{http://www.shanghairanking.com/}}. This survey lists the top 500 ranked
universities in the world, which we shall simply refer to as the ARWU
500.  Table~\ref{tab:ARWU500} lists the top 6 countries, as ranked by
the Academic Ranking of World Universities (ARWU) in 2012. The
national rankings are determined in a similar manner to those based
on the Olympic medal tallies. Countries are first ranked in descending
order by the number of university entries they have in the ARWU
Top 20, then by the number of Top 100 universities, followed by the
number of Top 200, 300, 400 and 500 entries respectively. %\footnote{\url{http://www.shanghairanking.com/ARWU-Statistics-2012.html#2}}

\begin{table}[!hb]
\centering
\begin{tabular}{lrrrrrr}
\sc Country & Top 20 & Top 100 & Top 200 & Top 300 & Top 400 & Top 500 \\
\hline
USA & 17 & 53 & 85 & 109 & 137 & 150 \\
UK & 2 & 9 & 19 & 30 & 33 & 38 \\
Japan & 1 & 4 & 9 & 9 & 16 & 21 \\
Australia & -- & 5 & 7 & 9 & 16 & 19 \\
Germany & -- & 4 & 14 & 24 & 30 & 37 \\
Canada & -- & 4 & 7 & 17 & 18 & 22 \\
\end{tabular}
\caption{\label{tab:ARWU500} Top 6 countries, as ranked by the Academic Ranking of World Universities (ARWU) in 2012.}
\end{table}

These 6 countries will form the basis of our study of the HPC
resources to available to academics in Australia, in comparison to our
overseas competitors. The list includes Japan, Germany and the USA,
the traditional leaders of the supercomputing field. Canada has
broadly similar socioeconomic characteristics to Australia and hence
provides a useful point of comparison.

We now turn our attention to the June 2012 Top 500 Supercomputer
list. 
%We will refer to the survey performed in June 2012. %\footnote{\url{http://top500.org/lists/2012/06}}.
%\footnote{\url{http://top500.org/lists/2012/06}} 
We filter the Top 500 supercomputing data by restricting ourselves to
the aforementioned top 6 countries in the ARWU ranking.  The top 3
entries for each country in the Academic and Research segments of the
Top 500 supercomputer list are displayed in
Table~\ref{tab:top500}. Also shown are the total number of entries,
number of compute cores, and combined computing
power for all Academic/Research entries in the list for that
country. The quantity that we will be interested in is the combined
$R_{\rm max}$ value for each country, which is an indicator of the
total number of Teraflops available to the Academic/Research segments
in that country. $R_{\rm max}$ is the LINPACK benchmark and provides a
measure of the supercomputer's speed in Teraflops.

\begin{table}[!tb]
\centering
\begin{tabular}{lrrrrr}
\textsc{Rank} & \textsc{Country/Site} & $N_{\rm cores}$ & $R_{\rm max}$ & $R_{\rm peak}$ \\ 
\hline\\[-3mm]
& \multicolumn{2}{l}{\textbf{\large Australia}} & & & \\
\hline
31 & VLSCI/Avoca & 65536 & 690.2 & 838.9 \\
139 & NCI-NF/Vayu & 11936 & 126.4 & 139.9 \\ %Sun Blade x6048, Xeon X5570 2.93 Ghz, Infiniband QDR	National Computational Infrastructure National Facility (NCI-NF)
248 & iVEC & 9600 & 87.2 & 107.5 \\ %HP Pod BL2x220, X5660 2.8 Ghz, Infiniband QDR	iVEC
& \bf Total: 3 Academic/Research entries & \bf 87072 & \bf 903.8 & \bf 1086.3 \\
\hline\\[-3mm]
& \multicolumn{2}{l}{\textbf{\large Canada}} & & & \\
\hline
66 & SciNet/U. Toronto/Compute Canada/GPC & 30912 & 261.6 & 312.82 \\ %GPC	xSeries iDataPlex, Xeon E5540 4C 2.53GHz, Infiniband	SciNet/University of Toronto/Compute Canada
71 &  Calcul Canada/Calcul Qu\'ebec/Sherbrooke & 37728 & 240.3 & 316.9 \\ %Rackable C2112-4G3 Cluster, Opteron 12 Core 2.10 GHz, Infiniband QDR	Calcul Canada/Calcul Québec/Université de Sherbrooke
90 & Environment Canada & 8192 & 185.1 & 251.4 \\ %	Power 775, POWER7 8C 3.84 GHz, Custom	Environment Canada
& \bf Total: 9 Academic/Research entries & \bf 137872 & \bf 1342.5 & \bf 1751.3 \\
\hline\\[-3mm]
& \multicolumn{2}{l}{\textbf{\large Germany}} & & & \\
\hline
4 & Leibniz Rechenzentrum/SuperMUC & 147456 & 2897.0 & 3185.1 \\ %iDataPlex DX360M4, Xeon E5-2680 8C 2.70GHz, Infiniband FDR	Leibniz Rechenzentrum
8 & Forschungszentrum Juelich/JuQUEEN &  131072 & 1380.4 & 1677.7 \\ %JuQUEEN	BlueGene/Q, Power BQC 16C 1.60GHz, Custom	Forschungszentrum Juelich (FZJ)
25 & Forschungszentrum Juelich/JUGENE & 294912 & 825.5 & 1002.7 \\ %JUGENE	Blue Gene/P Solution	Forschungszentrum Juelich (FZJ)
& \bf Total: 16 Academic/Research entries & \bf 753944 & \bf 7062.6 & \bf 8471.0 \\
\hline\\[-3mm]
& \multicolumn{2}{l}{\textbf{\large Japan}} & & & \\
\hline
2 & RIKEN/K computer & 705024 & 10510.0 & 11280.4 \\ %	K computer, SPARC64 VIIIfx 2.0GHz, Tofu interconnect	RIKEN Advanced Institute for Computational Science (AICS)
12 & IFERC/Helios & 70560 & 1237.0 & 1524.1 \\ %Bullx B510, Xeon E5-2680 8C 2.700GHz, Infiniband QDR	International Fusion Energy Research Centre (IFERC), EU(F4E) - Japan Broader Approach collaboration
14 & GSIC/Tokyo Inst. of Tech./TSUBAME 2.0 & 73278 & 1192.0 & 2287.6 \\ %HP ProLiant SL390s G7 Xeon 6C X5670, Nvidia GPU, Linux/Windows	GSIC Center, Tokyo Institute of Technology
& \bf Total: 23 Academic/Research entries & \bf 1184258 & \bf 17089.0 & \bf 20430.9 \\
\hline\\[-3mm]
& \multicolumn{2}{l}{\textbf{\large United Kingdom}} & & & \\
\hline
13 & STFC/Daresbury Laboratory/Blue Joule & 114688 & 1207.8 & 1468.0 \\ %Blue Joule	BlueGene/Q, Power BQC 16C 1.60GHz, Custom	Science and Technology Facilities Council - Daresbury Laboratory
20 & U. Edinburgh/DiRAC & 98304 & 1035.3 & 1258.3 \\ %BlueGene/Q, Power BQC 16C 1.60GHz, Custom	University of Edinburgh
32 & U. Edinburgh/HECToR & 90112 & 660.2 & 829.0 \\ %Cray XE6, Opteron 6276 16C 2.30 GHz, Cray Gemini interconnect	University of Edinburgh
& \bf Total: 16 Academic/Research entries & \bf 455584 & \bf 5875.3 & \bf 7553.0 \\
\hline\\[-3mm]
& \multicolumn{2}{l}{\textbf{\large United States}} & & & \\
\hline
1 & DOE/NNSA/LLNL/Sequoia & 1572864 & 16324.8 & 20132.7 \\ %Sequoia	BlueGene/Q, Power BQC 16C 1.60 GHz, Custom	DOE/NNSA/LLNL
3 & DOE/SC/Argonne/Mira & 786432 & 8162.4 & 10066.3 \\ %Mira	BlueGene/Q, Power BQC 16C 1.60GHz, Custom	DOE/SC/Argonne National Laboratory
6 & DOE/SC/Oak Ridge/Jaguar & 298592 & 1941.0 & 2627.6 \\ %Jaguar	Cray XK6, Opteron 6274 16C 2.200GHz, Cray Gemini interconnect, NVIDIA 2090	DOE/SC/Oak Ridge National Laboratory
& \bf Total: 87 Academic/Research entries & \bf 5063813 & \bf 44953.9 & \bf 56928.4 \\
\hline
\end{tabular}
\caption{\label{tab:top500}Selected entries in the June 2012 Top 500
  Supercomputer list in the Academic and Research segments. The top 3
  entries are listed for each of the chosen countries, as well as the
  total number of entries and the aggregate computing capacity of the
  entries. $N_{\rm cores}$ is the number of compute cores. $R_{\rm max}$ (the LINPACK benchmark score) and
  $R_{\rm peak}$ (the theoretical peak) are in Teraflops.}
\end{table}

The most straightforward measure of the supercomputing power available
to researchers in a given country would be to compare the integrated
$R_{\rm max}$ values in the academic segment. However, this simple
measure doesn't reflect the level of competition for those resources.
In order to provide a better estimate of the HPC resources available
to a given research group, we propose a novel measure called the
500/500, which is calculated for each country by taking the combined
Teraflops of the Academic and Research entries in the Top 500 supercomputer list
and dividing by the number of institutions in the ARWU 500. A summary
of the data is presented in Table~\ref{tab:500500}. The measure
assumes that the number universities in the ARWU 500 is a good
representation of the number of academic supercomputing groups in the
country.

\begin{table}[!h]
\centering
\begin{tabular}{lrrrr}
\sc Country & \sc Top 500 & \sc Total $R_{\rm max}$ & \sc ARWU 500 & 500/500 \\
\hline
Australia & 3 & 903.8 & 19 & 47.6 \\
Canada & 9 & 1342.5 & 22 & 61.0 \\
Germany & 16 & 7062.6 & 37 & 190.9 \\
Japan & 23 & 17089.0 & 21 & 813.8 \\
UK & 16 & 5875.3 & 38 & 154.6 \\
USA & 87 & 44953.9 & 150 & 299.7
\end{tabular}
\caption{\label{tab:500500} Data of interest for the selected countries in 2012. Listed are the number of Academic/Research Top 500 entries, the combined $R_{\max}$ (in Teraflops) under the Academic and Research segments, the number of ARWU 500 entries and our proposed 500/500 measure of academic HPC resources (in Tflops/institution).}
\end{table}

\begin{figure}[!h]
\centering
\tikzset{external/export next=false} % Need this for some reason?
\begin{tikzpicture}
\begin{axis}[
    ylabel=500/500 Score,
    symbolic x coords={Australia,Canada,Japan,Germany, UK, USA},
    xtick=data]
    \addplot[ybar,fill=blue] coordinates {
(Australia,	47.6)
(Canada,	61.0)
(Germany,      190.9)
(Japan, 813.8)
(UK,	154.6)
(USA,	299.7)};
\end{axis}
\end{tikzpicture}
\caption{\label{fig:500500} The 500/500 scores (in units of Tflops/institution) for the selected countries in 2012.}
\end{figure}
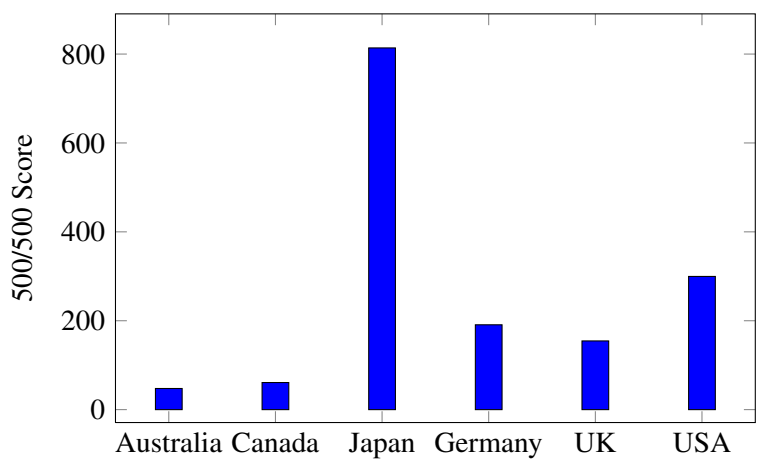

As demonstrated in Figure~\ref{fig:500500}, Japanese researchers are
the clear winners, with 500/500 score more than double that of
second-placed USA, and nearly twenty times that of Australia! While the
USA easily has the highest integrated $R_{\rm max}$ score, they are
ranked second on the basis of their 500/500 score, a reflection of the
intense competition for those resources as indicated by their place at
the top of the ARWU 500. Of the six selected countries, 
Australia ranks last according to the 500/500 metric.

%These results correlate well with the anecdotal experience of the
%author in the research field of Lattice QCD.

% giving some encouragment as to the
%validity of the 500/500 measure as a means of comparison.

% which provides supporting
%evidence for the 

%will that we
%will compare to are the top 5 ranked countries on the most recent
%Academic Ranking of World Universities (ARWU) list conducted by
%Shanghai Jiao Tong University, China.

%This survey lists
%the top 500 ranked universities in the world, which we shall refer to
%as the ARWU 500\footnote{\tt
%  http://www.shanghairanking.com/ARWU2011.html}.  On this survey
%Australia ranked $6^{\rm th}$ behind the USA, UK, Germany, Japan and
%Canada (refer Table~\ref{tab:500500}).

\section{GPU Computing}

As demonstrated in the previous section, Australian researchers are
disadvantaged with regard to HPC resources when compared to our overseas
competitors.  The lack of HPC resources is particularly acute in our field of Lattice QCD, where
 some of our competitors have access to dedicated lattice machines capable of hundreds of Teraflops. 
% and noting that we get a share of the order of a few percent of the major facilities while 
%Looking at the data above, the message is clear: our competitors in the US, Germany and
%Japan have access to many more times the cycles than we do.

%In fact, if not for the gauge configuration sharing program put in
%place by the ILDG we simply would not be able to compete at all. 
The ILDG program allows for the sharing of gauge field configurations
within a group or with the lattice QCD community at
large\cite{ildg}. The PACS-CS collaboration in Japan generously
released several gauge field ensembles of large volume and light quark
mass suitable for cutting edge calculations to the general lattice
community\cite{pacs-cs}. Through the use of these configurations we
have been able to bypass the unaffordable gauge field generation
process and devote our limited cycles towards the production of quark
propagators.

It should come as no surprise that with the relatively scarce level of
HPC resources available to us when compared to our competitors, we
have turned to GPUs as a cost-effective way of competing
with overseas groups. Lattice QCD has a geometric parallelism that
makes it ideally suited to be put on GPUs\cite{Egri:2006zm,Clark:2009wm}. 
%Typical (four-dimensional)
%lattice sizes for modern computations are of the order $32^3\times
%64,$ that is in excess of 2 million lattice sites. Each of these sites
%can be allocated a processing thread and hence one can make use of the
%massive parallelism offered by GPUs.
NVIDIA has two distinct GPU product lines that are relevant to
HPC. The Tesla line of cards specifically targets HPC users, whereas
the commodity GeForce graphics cards target the much bigger computer
gaming market. The specific cards that we are interested in are listed
in Table~\ref{tab:GPUs}. As we can see, in comparison to the GTX
cards, the Tesla GPUs feature improved double precision performance
and ECC memory. These features come at a cost however, with a Tesla
card costing roughly 4 times as much as a top-end GTX card.

\begin{table}
\centering
\begin{tabular}{cccccc}
\emph{Architecture} & \emph{GPU} & \emph{Cores} & \emph{Peak (SP)}  & \emph{Peak (DP)} & \emph{ECC} \\  
\hline
Fermi & GTX 580 & 512 & 1581 Gflops & 166 Gflops & No \\
& Tesla M2090 & 512 & 1331 Gflops & 665 Gflops & Yes \\
Kepler & GTX 680 & 1536 & 3090 Gflops & 95 Gflops & No \\
 & Tesla K20 & 2496 & 3520 Gflops & 1170 Gflops & Yes %\\
%Xeon Phi & $>50$ (MIC) & ??? & $>1$ Tflop 
\end{tabular}
\caption{\label{tab:GPUs} Previous (Fermi) and current (Kepler) generation NVIDIA GPUs. Shown are the number of CUDA cores, the peak floating point performance in single and double precision, and the ECC memory capability.}
\end{table}

Fortunately, the numerical requirements for quark propagator
generation are much less strict than those for gauge field
generation. The need to preserve unitarity during gauge field
generation typically requires double precision, and as the generated
gauge fields are not easily ``checked'' one also requires ECC
memory. In contrast, for quark propagators the tolerance when
calculating the application of the fermion matrix inverse is typically
$\sim 10^{-5},$ which means single precision is
sufficient. Furthermore, the solution to the linear system 
is easily verified, avoiding the need for ECC memory. Hence, GTX cards
are perfectly viable for quark propagator calculation.

\section{Adventures in Single Precision}

To obtain the action of the the inverse fermion matrix $D^{-1}$ on a
vector we calculate the solution to the linear system
\begin{equation}
D\vec{x} = \mathbf{b}. \label{eq:linsys}
\end{equation}
As the fermion matrix $D$ is non-Hermitian the most common
algorithm for obtaining the solution is BiCGStab\cite{bicgstab} or
some variant thereof. In double precision BiCGStab usually converges to a
solution, even though the typical convergence is not smooth but rather
`spiky'.  However, in single precision we find that BiCGStab is
numerically unstable. When attempting to invert the fermion matrix on
large lattices and light quark masses BiCGStab frequently fails to
converge. To avoid this, we propose to use an algorithm that minimises
the residual and hence will converge smoothly.

The conjugate gradient (CG) algorithm\cite{cg} minimises the residual, but is
only applicable to cases where the matrix being inverted is Hermitian
positive-definite (Hpd). There are two simple ways to convert our
original problem into a form suitable for the CG algorithm. The first
is to simply multiply (\ref{eq:linsys}) by $D^\dagger$ to obtain the
CGNR form of the normal equations,
\begin{equation}
D^\dag D\vec{x} = D^\dag \vec{b}. \label{eq:cgnr}
\end{equation}
The second is to solve the CGNE form of the normal equations
\begin{equation}
D D^\dag \vec{x}' = \vec{b}, \label{eq:cgne}
\end{equation}
where the solution to the original equations is given by $\vec{x} = D^\dag \vec{x}'.$

When solving the CGNE form of the normal equations, the residual for
the normal form $|D D^\dag \vec{x}' - \vec{b}|$ and the
residual for the original form $|D \vec{x} - \vec{b}|$
coincide by construction, so when CGNE converges we have obtained the
solution to the original equation to the desired tolerance
$\delta_{\rm tol}.$ Furthermore, we find that even in single precision the estimated residual $|\vec{r}|$ and the true residual coincide for the CGNE process.

In double precision, when the CGNR process converges this usually
implies that we have obtained the desired solution. However, in single
precision, the solution to (\ref{eq:cgnr}) converges well before we have obtained the
solution to (\ref{eq:linsys}). To work around this, we propose a
simple modification of the CGNR process. When the CGNR normal equation
converges with tolerance $\delta_{\rm ne},$ check if we have a
solution to the original equation within $\delta_{\rm tol}.$ If not,
adjust $\delta_{\rm ne}$ and restart CGNR with the current
solution. Our modified CGNR algorithm with restarts is presented in
Figure~\ref{fig:CGNR}.

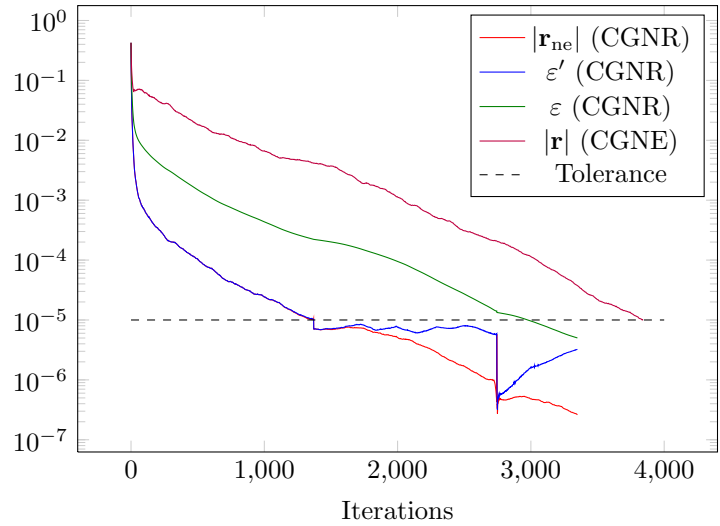
\begin{figure}[!tb]
\centering
\tikzsetnextfilename{combk13770}
\begin{tikzpicture}[scale=1.0]
\begin{semilogyaxis}[
xlabel=Iterations]
\pgfplotstableread{cgnrk13770.dat}\cgnrdata
\addplot[color=red,mark=none] table[y index=1] {\cgnrdata};
\addplot[color=blue,mark=none] table[y index=2] {\cgnrdata};
\addplot[color=dark-green,mark=none] table[y index=3] {\cgnrdata};
\pgfplotstableread{cgnek13770.dat}\cgnedata
\addplot[color=purple,mark=none] table[y index=1] {\cgnedata};
\addplot[color=black,domain=0:4000,dashed] {(1.0e-5)};
\legend{$|\vec{r}_{\rm ne}|$ (CGNR), $\varepsilon'$ (CGNR),  $\varepsilon$ (CGNR), $|\vec{r}|$ (CGNE), Tolerance}
\end{semilogyaxis}
\end{tikzpicture}
\caption{\label{fig:combk13770} Typical behaviour of the CGNE process and the CGNR process with restarts. Shown for CGNR are the estimated residual $|\vec{r}_{\rm ne}|$ and true residual  $\varepsilon'$ for the normal equation, as well as the true residual $\varepsilon$ for the original equation. For CGNE we show the estimated residual $|\vec{r}|$ (which coincides with the true residual).}
\end{figure}

A comparison of the typical behaviour of CGNE and our CGNR with
restarts is shown in Figure~\ref{fig:combk13770}. We can see that the
estimated residual $|\vec{r}_{\rm ne}|$ and the true residual for the
CGNR normal equations $\epsilon' = |D^\dag (D\vec{x} - \vec{b})|$
coincide until the CGNR system (\ref{eq:cgnr}) has converged, after
which they diverge due to hitting the limits of single precision. What
is interesting is that even though the CGNR process undergoes
restarts, the true residual for the original system $\epsilon =
|D\vec{x} - \vec{b}|$ decreases smoothly until it has converged to
within the desired tolerance.  Tests comparing CGNR and CGNE were
performed at several quark masses and we found that the modified CGNR process
(with restarts) consistently converges significantly faster than CGNE, requiring $\sim
10\%-30\%$ less iterations to reach the desired tolerance.

\begin{figure}
\hrule 
\vspace{6pt}
\begin{algorithmic}
\STATE Initialise $\delta_{\rm ne} := \delta_{\rm tol}$ to the desired solution tolerance.
\LOOP
\STATE Set $\vec{y} := \vec{r}_{\rm ne} := D^\dag D\vec{x} - D^\dag \vec{b},\quad \rho := |\vec{r}_{\rm ne}|^2.$
\WHILE{$\sqrt{\rho} > \delta_{\rm ne}$}
\STATE Set $\beta := \inner{\vec{y}}{D^\dag D\vec{y}},\quad \omega := \rho/\beta.$
\STATE Set $\vec{x} := \vec{x} + \omega\vec{y},\quad \vec{r}_{\rm ne} := \vec{r}_{\rm ne} - \omega D^\dag D\vec{y}.$
\STATE Set $\rho' := \rho,\quad \rho := |\vec{r}_{\rm ne}|^2,\quad \theta := -\rho/\rho'.$ 
\STATE Set $\vec{y} := \vec{r}_{\rm ne} - \theta\vec{y}.$
\ENDWHILE
\STATE Set $\epsilon := |D\vec{x} - \vec{b}|$ to the true residual for the original equation.
\STATE \textbf{if} $\epsilon < \delta_{\rm tol}$ \textbf{then exit} \COMMENT{\emph{We are finished.}}
\STATE Set $\epsilon' := | D^\dag D\vec{x} - D^\dag \vec{b}|$ to the true residual for the normal equation. 
\STATE Update  $\delta_{\rm ne} := \tau \cdot \delta_{\rm tol} \cdot (\epsilon'/\epsilon). $ \COMMENT{\emph{Restart CGNR.}} 
\ENDLOOP
\end{algorithmic}
\vspace{4pt}
\hrule
\caption{\label{fig:CGNR} The modified CGNR algorithm with restarts. The constant $\tau \sim 0.9$ controls the restart frequency.}
\end{figure}

%\section{...}

%\bibliographystyle{jhep}
%\bibliography{reference}

\providecommand{\href}[2]{#2}\begingroup\raggedright\endgroup

\end{document}